\documentclass[useAMS,usegraphicx,usenatbib]{mn2e}
\usepackage{graphicx, color}
\usepackage{aas_macros}     
\usepackage{amsmath}

\usepackage{fixltx2e}[1999/12/01]

\newcommand{\Msun}{ \ensuremath{M_{\odot}} }

\title[The journey of QSO haloes from z=6 to the present]
{The journey of QSO haloes from $z\sim6$ to the present}

\begin{document}
\setlength{\topmargin}{-1.cm}

\author[Angulo et al]{
\parbox[h]{\textwidth} {R. E. Angulo$^{1}$, V.~Springel$^{2,3}$,
  S.~D.~M.~White$^{1}$, S.~Cole$^4$, A.~Jenkins$^{4}$, C.~M.~Baugh$^{4}$,
  C.~S.~Frenk$^{4}$} \vspace*{6pt} 
\\ 
\\ $^1$ Max-Planck-Institute for Astrophysics, Karl-Schwarzschild-Str. 1, 85740 Garching, Germany.
\\ $^2$ Heidelberg Institute for Theoretical Studies, Schloss-Wolfsbrunnenweg
35, 69118, Heidelberg, Germany,
\\ $^3$ Zentrum f\"{u}r Astronomie der
Universit\"{a}t Heidelberg, ARI, M\"onchhofstr. 12-14, 69120 Heidelberg,
Germany.
\\ $^4$ Institute for Computational Cosmology, Dep.~of Physics,
Univ.~of Durham, South Road, Durham DH1 3LE, UK } \maketitle

\date{\today}
\pagerange{\pageref{firstpage}--\pageref{lastpage}} \pubyear{2011}
\label{firstpage}

\begin{abstract} 
We apply a recently developed scaling technique to the Millennium-XXL, one of the largest cosmological
N-body simulations carried out to date ($3\times 10^{11}$ particles
within a cube of volume $\sim 70\,{\rm Gpc^3}$). This allows us to investigate
the cosmological parameter dependence of the mass and evolution of haloes in
the extreme high-mass tail of the $z=6$ distribution. We assume these objects
to be likely hosts for the population of rare but ultraluminous high-redshift
quasars discovered by the Sloan Digital Sky Survey. Haloes with a similar
abundance to these quasars have a median mass of $9\times 10^{12}\,\Msun$ in
the currently preferred cosmology, but do not evolve into equally extreme
objects at $z=0$. Rather, their descendants span the full range conventionally
assigned to present-day clusters, $6\times10^{13}$ to $2.5\times
10^{15}\Msun$ for this same cosmology. The masses both at $z=6$ and at $z=0$
shift up or down by factors exceeding two if cosmological parameters are
pushed to the boundaries of the range discussed in published interpretations
of data from the WMAP satellite. The main factor determining the future growth
of a high-mass $z=6$ halo is the mean overdensity of its environment on scales
of $7$ to $14$~Mpc, and descendant masses can be predicted 6 to 8 times more
accurately if this density is known than if it is not. All these features are
not unique to extreme high-$z$ haloes, but are generic to hierarchical growth. 
Finally, we find that extreme haloes at $z=6$ typically acquired about half of 
their total mass in the preceding $100$~Myr, implying very large recent accretion 
rates which may be related to the large black hole masses and high luminosities 
of the SDSS quasars.
\end{abstract}

\section{Introduction}

Bright $z\sim6$ quasars are extremely rare objects. Their luminosity is thought
to be a result of supermassive black holes accreting gas at enormous rates
\citep{Fan2003,Kollmeier2006,Shen2008} at a time when only $\sim10\%$ of the
mass in the Universe was in haloes where gas could cool efficiently and about
half was still in diffuse form \citep{Angulo2010a}. If these quasars are
long-lived, then their dark matter haloes belong to an equally extreme tail
(presumably the most massive tail) of the halo distribution, a hypothesis
supported by the strong observed clustering of high-$z$ quasars
\citep[e.g.][]{Shen2007}. The observed QSO number density (about one per cubic
Gigaparsec) combined with abundance matching implies that their host halo
masses are well above $10^{12}\,\Msun$ at $z=6$. These masses correspond to
much more extreme peaks in the initial Gaussian density fluctuation field than
those associated with even the largest galaxy clusters today. Such quasars are
clearly exceptional events in a $\Lambda$CDM universe, but their properties,
together with those of the intergalactic absorption seen in their spectra,
encrypt key information about many astrophysical processes
\citep[e.g.][]{Fan2006NAR} and perhaps also about the background cosmological
model.

Studying the properties, environment and fate of the high-mass haloes in which
the $z\sim 6$ quasars may live is a challenging task for cosmological N-body
simulations. It requires a very large computational volume to obtain a
representative sample of extremely rare objects, sufficient spatial and mass
resolution to resolve their structure reliably, and sufficient time resolution
to build the merger trees needed to trace evolution throughout cosmic time.
Previous studies in this area have relied on analytic models calibrated using
simulations of less extreme objects \citep[e.g.][]{Trenti2008}, or resimulation
techniques applied to a limited number of systems
\citep{Li2007,Sijacki2009,Romano-Diaz2011}.  In this paper, we present an
extremely large N-body calculation, which meets the computational challenges
directly by simulating a very large volume at adequate resolution within the
$\Lambda$CDM paradigm.

With this simulation in hand, we study a variety of topics associated with the
assembly and the future of quasar haloes. For example, where are today's
descendants of the massive black holes that powered quasars at $z\sim6$?  This
question was explored by \cite{Springel2005a} using the Millennium Simulation
(MS). They concluded that these black holes would today lie at the centres of
cD galaxies in massive galaxy clusters. However, the MS is too small to contain
even one object like the SDSS $z\sim 6$ quasars, so this conclusion was based
on a small number of less rare systems. \cite{Trenti2008} used analytic tools
to extend these MS-based results, arguing for a much greater diversity in the
present-day descendants of SDSS QSO haloes.  A similar conclusion was reached
by \cite{DiMatteo2008} with cosmological SPH simulations.  Here, we are able to
use fully resolved N-body merger trees for haloes identified at the same space
density as the brightest SDSS quasars.  We confirm that these massive $z=6$
haloes should evolve into a variety of systems today. Most of their associated
black holes should end up in the central galaxies of haloes ranging from rich
groups to clusters. We show that the median mass of descendants approaches
$10^{15}\Msun$ but this value and that of early haloes themselves depend on
the parameters assumed for the background cosmological model.

Another issue we investigate is the large-scale environment surrounding
extreme high-redshift haloes. Such haloes are expected to be strongly biased
towards overdense regions. Although we confirm this, we also show that there
is considerable scatter, and many extreme haloes have environments of moderate
overdensity. Indeed, some even have environments which are slightly
underdense.

Finally, we explore the assembly histories predicted for extreme $z=6$
haloes for three different assumptions about the parameters underlying the
background $\Lambda$CDM cosmology. We measure the accretion rates onto these
haloes over the time period immediately preceding the epoch of observation and
show that these can be extremely high. This may well be related to the
fuelling of the extraordinarily high luminosities and masses measured for the
$z\sim 6$ quasars.

This paper is set out as follows. In Section 2, we present technical details
of our simulation and of the methods we use to identify dark matter haloes and
their evolutionary paths. Section 3 then presents our results, including the
properties of the $z=0$ descendants of high-$z$ quasar haloes, and their
assembly histories at higher redshift.  Our conclusions are presented in
Section 4.

\section{Numerical Methods}

In this section we describe the numerical tools used in this paper. We first
present our N-body simulation (Section~2.1), including a description of the
construction of halo and subhalo catalogues and of merger trees (Section~2.1.1). In
Section~2.2 we then explain how these numerical data can be used to explore
structure formation in cosmologies other than the one used to carry out the
original simulation. The last subsection (Section~2.3) defines the samples of
extreme haloes that we will study in the remainder of the paper.

\subsection{The MXXL N-body simulation}

\begin{figure*} 
\includegraphics[width=16cm]{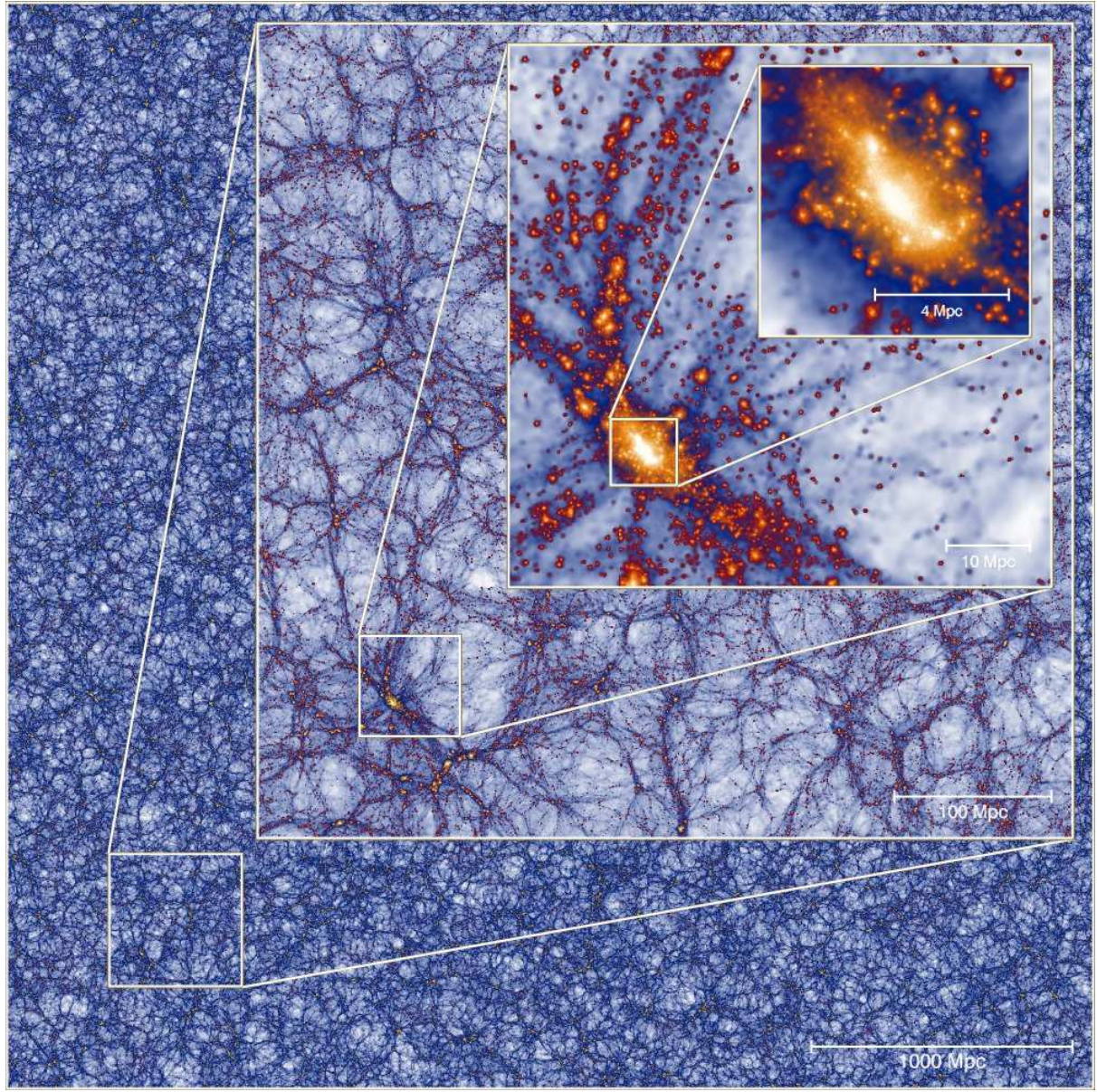} 
\caption{ Images of the matter density field in the Millennium-XXL focusing on
  the most massive halo present in the simulation at $z=0$. Each inset zooms
  by a factor of $8$ from the previous one; the side-length varies from
  4.1~Gpc down to 8.1~Mpc.  The intensity of each pixel is proportional to the
  logarithm of the dark matter density projected through a $25$~Mpc thick
  slab. This simulation has a dynamic range of $10^5$ on each spatial
  dimension, simultaneously resolving the internal structure of collapsed
  objects and the large-scale quasi-linear fluctuations in a $\Lambda$CDM
  universe.
\label{fig:picture}} 
\end{figure*}

Our simulation, named the Millennium-XXL or MXXL, represents the matter
content of a $\Lambda$CDM universe using more than $303$ billion particles
($6720^3$) within a comoving cube of side 4.1~Gpc.  Our choice of cosmological
parameters is identical to that used in the other Millennium Simulations
\citep[see Table~1,][]{Springel2005a, Boylan-Kolchin2009}, implying a particle
mass of $8.45\times10^9\Msun$.  This set of parameters is inconsistent with
the most recent constraints from observations of the Cosmic Microwave
Background and low-redshift large-scale structure \citep{Komatsu2011}, but, as
discussed in Section~2.2, our results can be scaled accurately to any nearby
cosmology, including those that are now more favoured. The comoving softening
length of the simulation was $\epsilon = 13.7$~kpc, implying $\sim10^{15}$
effective resolution elements in the full simulated volume  (the forces
are exactly Newtoninan beyond $2.8\times\epsilon$).  The enormous
statistical power and dynamical range that this implies is illustrated in
Fig.~\ref{fig:picture}, which shows the $z=0$ density field on four different
scales, starting from the whole box in the background, then zooming
progressively onto the most massive halo in the insets.

The initial phase-space distribution of the particles was set up at $z=63$ by
perturbing a glass-like distribution \citep{White1996,Baugh1995} using second
order Lagrangian perturbation theory \citep[2LPT][]{Scoccimarro1998}. The use of
2LPT has several advantages over the more common Zeldovich approximation
\citep{Zeldovich1970}, including small and rapidly-decaying transients in the
matter clustering and a better representation of the perturbations that seed
extremely large dark matter haloes \citep{Crocce2006, Knebe2009, Jenkins2010}.
The latter is particularly relevant for this paper. The amplitude of
individual Fourier modes was set hierarchically as described in 
Jenkins~(2012, in prep). This allows an efficient and consistent generation of initial
conditions for the resimulation of any selected subregion at, in principle,
arbitrarily high mass resolution.

Note that the MXXL is the first simulation with a large enough volume and a
small enough particle mass to contain a representative sample of well-resolved
\citep[$\ge 1000$ particles,][]{Trenti2010} haloes which could represent the
hosts of the $z\sim 6$ SDSS quasars.  Performing a simulation with these
characteristics posed severe computational challenges with respect to raw
execution time, scalability of the simulation algorithms, memory consumption and
I/O performance. Thanks to a highly optimised simulation code and one of the
largest supercomputers in Europe, these challenges were successfully met.
Specifically, the MXXL was carried out with a special version of the {\tt
GADGET-3} code \citep{Springel2005b}, which aggressively reduced peak memory
consumption at runtime, incorporated a number of analysis tools on the fly, and
implemented a special compression of the output data. As a result the MXXL was
completed in late summer 2010 in less than 3 million CPU hours (including
postprocessing calculations), using 30~Tb of RAM, and $12,228$~cores of the
Juropa cluster at the J\"ulich Supercomputer Center in Germany. We refer to
\citet{Angulo2012} for more details about the simulation.

\subsubsection{Halo and Subhalo catalogues}

We identified self-bound halo/subhalo structures throughout the MXXL at the
same 64 redshifts used in the MS and MS-II.  This output frequency (roughly
equally spaced in time by $300$ Myr for $z < 2$, and by $100$ Myr at $z\sim6$)
allows us to build detailed merger trees.  At each output time, we first apply
a Friends-of-Friends (FoF) algorithm \citep{Davis1985}, with a linking length
of $0.2$ times the mean interparticle separation to build a FoF group
catalogue down to a limit of 20 particles. We then use a memory-efficient
implementation of the {\tt SUBFIND} algorithm \citep{Springel2001b} to
identify self-bound substructures within each FoF group down to a limit of
$15$ particles. These calculations were performed {\it on-the-fly}
during the N-body calculation, so that it was not necessary to store the
particle data at all output times. This significantly reduced the I/O and
storage requirements of the simulation.

Summing over all output times, there are $2.5\times 10^{10}$ FoF groups in the
MXXL with more than $20$ particles. At $z=6$ there are $3.7\times 10^7$ such
groups and $6.5\times 10^8$ at $z=0$. The most massive FoF group at $z\sim6$
contains 3,285 particles and $4$ substructures. This is about $300$ times
less massive than the biggest halo at the $z=0$ snapshot, which contains 
1,062,232 particles and $688$ substructures with more than $15$ particles.

Finally we built ``merger trees" similar to those described in
\cite{Springel2005a} in order to follow the evolution of halo/subhalo
structure in detail. For every subhalo in our catalogues we define a pointer
to a unique descendant in the subsequent snapshot by locating the subhalo
containing the greatest number of its $15$ most bound particles\footnote{If
two subhaloes contain the same number of these particles, we choose 
the one with the largest total binding energy.}. These
pointers are used to create a tree-like data structure, which represents the
full assembly history, the current substructure, and the future evolution of
every halo. In particular this allows us to map out the formation histories of
our putative $z=6$ quasar hosts, and to follow their later evolution down to
$z=0$.

\subsection{Exploring structure formation in other cosmologies}
\begin{small}
\begin{table}
\begin{center}
\begin{tabular}{llllll}
\hline
\hline
&  $m_{\rm dm}$ & Box & $\Omega_{\rm m} $ & $\Omega_{\rm b}$ & $\sigma_8$  \\
\hline
\tt{MXXL(M1)} & $0.95$ & $4167$ & $0.250$ & $0.045$ & 0.9  \\
\tt{M3}       & $1.64$ & $5091$ & $0.238$ & $0.0416$  & 0.761 \\
\tt{M7}       & $1.26$ & $4467$ & $0.272$ & $0.0416$ & 0.807 \\
\hline
\end{tabular}
\end{center}
\caption{Parameters of the original MXXL simulation and of the scaled versions
  used to represent other cosmologies.  The columns are as follows: (1) the
  name of the simulation; (2) the mass of a dark matter particle in units of
  $10^{10}\,\Msun$; (3) the side of the computational box in units 
  of Mpc; (4) the total matter density; (5) the baryon density; 
  (6) The linear fluctuation amplitude at $z=0$. In all
  cases the primordial spectral index is $n_s = 1$, the Hubble constant at
  $z=0$ is $H_0 = 73$~km/s/Mpc, and the dark energy is assumed to be a
  cosmological constant.}
\label{tab:params}
\end{table}
\end{small}

The MXXL was an extremely expensive numerical simulation and so could only be
carried out once for a specific set of cosmological parameters. However, the
rescaling method of \cite{Angulo2010b} allows us to use the simulation to
analyse any neighbouring cosmology with Gaussian initial fluctuations, and in this paper we will show
results not only for the original MS cosmology but also for two other versions
of the standard $\Lambda$CDM cosmology. The accuracy of the rescaling scheme is
remarkably high if it is applied carefully; masses of individual objects are
reproduced to better than 10\% and positions to better than 100~kpc
\citep{Angulo2010b,Ruiz2011}. In the following, we briefly recap the main
features of the method.

First, consider a ``target'' cosmological model at $z=z_B$ which we seek to
match using the results of an ``original" cosmological simulation of
side-length $L_A$ (in units of $h^{-1}$~Mpc).  The heart of the method is to find a length transformation,
$L_A \rightarrow  L_B = s~L_A$, and a relabelling of the time variable $z_A
\rightarrow z_B$, by requiring that the variance of the linear density field in
the target cosmology, $\sigma_B^2(R,z_B)$, over the range $[R_1, R_2]$ is as
close as possible to that of the original cosmology, $\sigma_A^2(R, z_A)$, over the
range $[s^{-1}R_1,s^{-1}R_2]$ at redshift $z_A$.  Thus, we minimise:

\begin{equation}
\int_{R_1}^{R_2} \frac{{\rm d}R}{R} \left[ \sigma_B^2(R)D_B(z_B) - \sigma_A^2(s^{-1}R)D_A(z_A) \right]^2,
\label{eq:rms}
\end{equation}

\noindent over $s$ and $z_A$, where $D(z)$ is the linear growth factor in units
of its present-day value.

In the Press-Schechter theory \citep{Press1974} the halo mass function is 
determined by the
linear variance of the underlying dark matter field, thus Eq.~(\ref{eq:rms})
also minimises the difference between the halo mass functions in the
target and original cosmologies over the mass range $[M(R_2),M(R_1)]$. 
We usually take $M(R_2)$ to be the mass of the largest halo in the simulation 
at the lowest redshift of interest ($z=0$ here) and $M(R_1)$ to be that of
the least massive resolved halo.

As a result, the original box size will be expanded by a factor $s$, and the
output at redshift $z_A$ will represent redshift $z_B$ in the target
cosmology. Redshifts in the target cosmology ($z'$) corresponding to the
redshifts ($z$) of stored data in the original simulation are then
determined implicitly by $D_B(z') = [D_A(z) / D_A(z_A)] D_B(z_B)$. The
mass of a simulation particle (in units of $\Msun$) in the target cosmology is $m_B =
(\Omega_{m,B}H_B^2/\Omega_{m,A}H_A^2)\,s^3 \,m_A$, where $\Omega_{m,X}$ and
$H_X$ ($X=A$ or $B$) are the dimensionless total matter densities and Hubble
parameters of the two cosmologies.

The second part of the algorithm of \cite{Angulo2010b} corrects differences in
power spectrum shape on large scales between the original and target
cosmologies by altering the amplitude of quasi-linear modes using the
Zel'dovich approximation.  In this paper we do not make this correction since
we are interested in the internal structure, abundance and evolution of
massive haloes rather than in their spatial distribution, and
we use only the original cosmology when looking at the overdensity around
haloes (a quantity that is slightly affected by our scaling). 

With this technique we have created two additional halo catalogues in
alternative cosmologies which we denote M3 and M7. These have cosmological
parameters motivated by the 3-year \citep{spergel2007} and 7-year
\citep{Komatsu2011} analyses of data from the WMAP satellite.  The main feature
of these models are values for $\sigma_8$ which are lower than in the MXXL and
the other Millennium Simulations, and different values for $\Omega_m$ (see
Table~1).  The corresponding length scalings are $s=1.222$ and $s=1.072$ for M3
and M7, respectively.  The $z=0.623$ and $z=0.319$ outputs of the MXXL
represent $z'=0$ in the M3 and M7 cosmologies.

\subsection{QSO haloes}
\label{sec:qso}

In this paper we will assume that the QSO luminosity is a monotonically-increasing
function of the FoF host halo mass at any given time, and that there is a duty
cycle that is independent of the halo mass. Models with these characteristics 
appear to be preferred by clustering analyses \citep[e.g][]{White2008,Bonoli2010}, but 
we note that they are not the only possibility. In physically motivated models the 
QSO host depends on the details of the galaxy formation model and in particular on 
AGN feedback \citep{Marulli2008,Bonoli2009,Fanidakis2011,Fanidakis2012}. 

Therefore, we consider halo samples limited by FoF mass at two different thresholds
corresponding to two different abundances at $z\sim 6$ which we keep constant
across our three cosmologies, ``Long-lived QSO haloes'' (LLQ haloes) have a
comoving number density of $n=0.4\,{\rm Gpc}^{-3}$, requiring FoF masses
above $[15.4,\,9,\,5]\times10^{12}\,\Msun$ for the M1, M7 and M3 cosmologies, 
respectively. The number density of this sample (which is well below the limit 
that could be probed by the MS) matches that of the extremely bright 
quasars observed at $z>5$ in the SDSS \citep[$n=0.6\,{\rm Gpc}^{-3}$,
][]{Fan2003,Fan2006}. Thus, they mimic the assembly history of SDSS QSOs if they
have a $100\%$ duty cycle, i.e. if they shine constantly at their full brightness.

Our second sample, ``Short lived QSO haloes'' (SLQ haloes) is selected at 30
times higher abundance, i.e.  $n=11.6\,{\rm Gpc}^{-3}$.  This corresponds to
minimum FoF halo masses of $[7.0,\,4.2,\,2.0]\times10^{12}\,\Msun$ for the M1, M7
and M3 cosmologies.  This sample could be regarded as representing the hosts of
the high-redshift SDSS quasars if these have a 1/30 duty cycle, i.e. each
object shines at full brightness only 1/30 of the time.

Due to our length scaling, the total number of haloes in our two samples varies
by factors of $s^3$ between the three cosmologies; the SLQ samples contain $810$,
$997$ and $1478$ objects for the M1, M7 and M3 cases, respectively (note that
we expect four objects above this mass threshold in the MS which, in fact,
contained only two), whereas the LLQ samples contain $27$, $34$, and $50$ haloes. The
redshift of the 
samples also differs between cosmologies because the MXXL data were stored only
at a discrete set of times: we use $z=6.18$ (M1), $z=6.19$ (M3) and $z=5.96$
(M7) for the three cases.

Finally, we note that alternative sample definitions, for example, using
virial masses or circular velocities rather than FoF masses, do not produce
significant differences in our results. $70\%$ of the $1000$ most massive FoF
haloes rank within the $1800$ most massive haloes according to $M_{200}$, and
within the $4000$ according to peak circular velocity.

\section{RESULTS AND DISCUSSION}

Using the numerical tools and halo samples presented in the previous section,
we now study the evolution of the host haloes of $z\sim6$ quasars. We look
into the type of objects they turn into at the present day (Section~3.1) and how the
masses of these descendants are related to the environments of their $z\sim 6$
progenitors (Section~3.3). In addition, we explore the growth rates of our quasar
host haloes both prior (Section~3.4) and subsequent (Section~3.2) to their identification
at $z\sim 6$.

\subsection{The mass and fate of $z\sim6$ quasar hosts}

\begin{figure} 
\includegraphics[width=8.5cm]{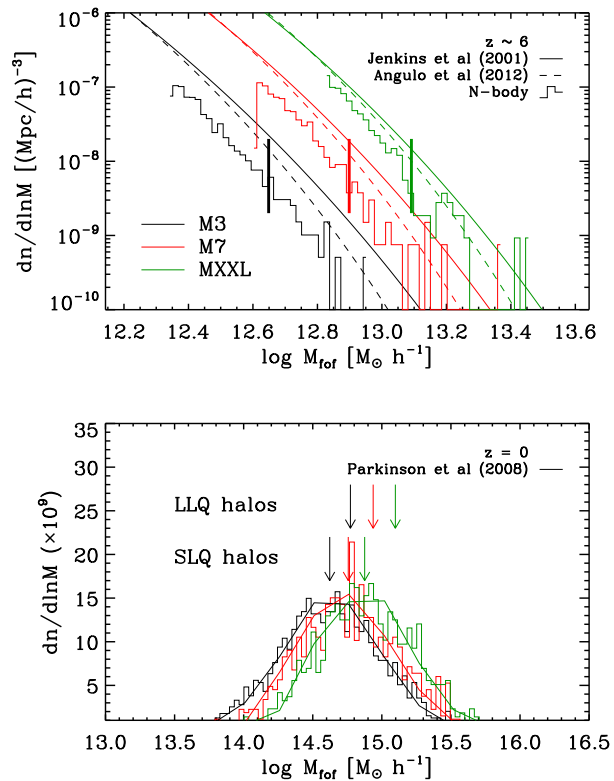} 
\caption{ Comparison of the FoF mass distributions of our SLQ host halo
  samples at $z\sim 6$ to those of their descendants at $z=0$. The top panel
  shows the differential number density of haloes for the three cosmologies of
  Table~1, corresponding to the parameters preferred by analyses of one, three
  and seven years of data from the WMAP satellite.  On each histogram a thick
  vertical line indicates the mass threshold for the corresponding LLQ host
  halo sample. The bottom panel shows the distributions of FoF masses
  of the $z=0$ descendants of these haloes. Predictions of the \citet{Jenkins2001}
  and \citet{Angulo2012} fitting formulae, and
  of merger trees constructed using the \citet{Parkinson2008} algorithm are
  shown in the top and bottom panels, respectively.  Vertical
  arrows in the bottom panel indicate the median masses of the descendants of
  SLQ haloes ($n=30\,{\rm Gpc}^{-3}$) and of LLQ haloes ($n=1\,{\rm
    Gpc}^{-3}$). Note that haloes of similar mass at $z\sim6$ end up in haloes
  with a wide range of masses at $z=0$.  
  \label{fig:mf} } 
\end{figure}

Since we are assuming that high-redshift quasars live in the most massive
objects present at that time, one might naively expect that their
descendants will lie at the centres of the most massive haloes at any later
time, in particular, in the central galaxies of large galaxy clusters today.
However, in this section we confirm the results of previous studies showing
this is not generally true \citep{Trenti2008, Overzier2009}. The descendants 
of QSOs can be found in haloes spanning a factor of $30$ in mass, and in a few 
cases, they are not even located in the central object of this halo, but in an 
orbiting subhalo.

The upper panel of Fig.~\ref{fig:mf} shows differential mass functions for the
two $z\sim 6$ QSO host halo samples (defined in Section~\ref{sec:qso})  and for
the three cosmological models of Table~1, which differ primarily in their value
of $\sigma_8$ and $\Omega_m$. By construction, these samples consist of very
massive haloes. Their mean mass ranges from $2.6$ to $8\times10^{12}\,\Msun$ for
the SLQ samples, and from $5.2$ to $15.2\times10^{12}\,\Msun$ for the sparser 
LLQ samples -- for both samples these mean masses increase smoothly with $\sigma_8$,
as expected. Within each sample more than than $99\%$ of all haloes lie
within a factor of two in mass of the threshold for inclusion. This is a 
consequence of the exponentially falling high-mass tail of the halo mass function.

The magnitude of the shifts between the three cosmologies are well described
by the \cite{Jenkins2001} and \cite{Angulo2012} fitting formulae, which we
display as solid or dashed lines in Fig.~\ref{fig:mf}. However, at this redshift these
formulae overpredict the number of haloes of a given mass by a factor of two
to three. In part, this reflects the fact that these formulae were calibrated using
simulations and redshifts where the most massive haloes were much less extreme 
than those considered here, but the disagreement might also be in part a consequence of
the non-universal behaviour of the halo mass function \citep[see e.g][]{Tinker2008}. 

We identify the $z=0$ descendant of each halo in our samples as the object
that contains the majority of its 15 most bound particles. 
Although in principle it is possible that a different structure contains most
of the mass of our high-$z$ haloes, following the innermost particles should
represent well the fate of a hypothetical black hole sitting at the centre of
the halo.  The bottom panel of Fig.~\ref{fig:mf} displays the FoF mass
distributions of these descendant haloes. For comparison, we also present
predictions made using the analytic merger tree algorithm of
\cite{Parkinson2008} which is based on the Extended Press-Schechter (EPS)
formalism and calibrated using merger trees extracted from the Millennium
Simulation.

The median FoF mass of the descendants of our SLQ haloes varies by a
factor of just $1.7$ across our three cosmologies (see the lower set of coloured
arrows in Fig.~\ref{fig:mf}). This is significantly smaller than the
spread of a factor of $3.1$ in the the initial median masses. The
typical descendant mass is $M\sim 5.6\times10^{14}\,\Msun$, corresponding
to a moderately rich $z=0$ cluster. The median masses of the descendants
of the sparser LLQ haloes are larger by a factor of about $1.5$ and
also vary slightly more with $\sigma_8$ (see the upper set of coloured arrows
in Fig.~\ref{fig:mf}). The median initial masses of the LLQ
haloes are typically a factor of $1.9$ larger than those of the SLQ haloes,
so both within a single cosmology and between cosmologies the evolution
is convergent in the sense that descendant masses are more similar than those
of the original $z\sim 6$ objects. This is probably a result of QSO haloes, in
all cases, descending into much less extreme peaks for which the
differences among cosmologies are smaller. For example, the mass function for
the M1 and M7 cosmologies are almost identical at $z=0$ for masses below $M \sim
10^{14}\,\Msun$.

In all cases, there is a large spread in the masses of the descendants, as
shown by the bottom panel of Fig.~\ref{fig:mf}. This distribution at $z=0$ can
be well described by a log normal with $\sigma = 0.36$~dex.  The most massive
tail indeed corresponds to rare, high-mass clusters, but the least massive one
corresponds to galaxy groups.  The scatter is slightly smaller for the LLQ
sample. We also would like to note that every descendant of an LLQ halo is the
dominant structure of its $z=0$ group, but in the M1 cosmology $39$ out of the
$810$ descendants of SLQ haloes are satellite subhaloes. For the M3 and M7
cases the corresponding numbers are similar.  As we will see in the next
subsection, all this reflects the variety of formation histories among dark
matter haloes of similar mass.  

Our distributions are in good qualitative agreement with the EPS-based
calculations of \cite{Trenti2008}. Using the same M1 cosmology, these authors
found $68\%$ of the descendants of a sample analogous to our LLQ to have masses
in the range $2.5$ to $12.2\times10^{14}\,\Msun$ with a median of
$5.6\times10^{14}\,\Msun$.  For this cosmology, our results show a similar 
scatter, but with a median which is $35\%$ larger. As
can be seen in Fig.~\ref{fig:mf} for all cosmologies, similar scatter and median 
masses are predicted by the merger tree algorithm of \cite{Parkinson2008} if
we use a halo sample that matches the number density of the SLQ haloes.
This confirms that our scaling algorithm captures the main features of
structure growth in different background cosmologies.

\subsection{The journey to $z=0$}

\begin{figure} 
\includegraphics[width=8.5cm]{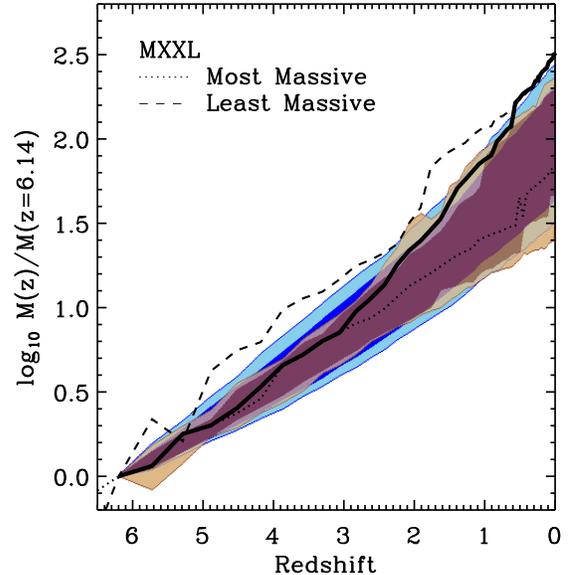} 
\caption{ Mass growth of high-redshift quasar host haloes in the M1 cosmology
  (i.e. in the original MXXL). Blue regions show mass growth for the SLQ
  sample, whereas brown regions show growth for the sparser and more massive
  LLQ sample. In each case, the dark and light regions enclose $68\%$ and
  $86\%$ of the trajectories. Individual lines highlight particular
  trajectories. The thick solid line is the growth history that the largest
  halo at $z=6$ would have if its mass was equal to that of the most massive
  halo in the simulation at all later redshifts. In contrast, the dotted line
  shows the actual growth history of this halo. Finally, the dashed line
  displays the growth of the lowest mass halo in the SLQ sample which, by
  chance, is one of the fastest growing of all haloes.  \label{fig:traj}}
\end{figure}

We now look in more detail at the paths connecting our samples of $z\sim 6$
quasar host haloes to their present-day descendants, concentrating on
results in the unscaled MXXL, i.e. in the original M1 cosmology.  In
Fig.~\ref{fig:traj} we plot mass growth, defined as the ratio of descendant
mass at each redshift to initial mass at $z\sim 6$.  Light blue and brown
regions indicate $86\%$ of the trajectories for SLQ and LLQ samples,
respectively. Darker regions of each colour outline the regions containing
68\% of the trajectories. Growth histories seem to be remarkably similar in
the two samples and to be faster than exponential, described approximately by
$\log_{10} M(z)/M(z=6.19) = 0.21\times (6.19-z)^{1.2}$. This seems independent
of initial mass at $z=6$, at least over the relatively restricted range of
(high) masses considered here. Of course, this formula only describes the
typical behaviour, and very different growth histories occur for haloes of
similar initial mass. Some massive $z=6$ haloes have grown only by a factor of
20 to 30 by $z=0$ -- much less than the change of the characteristic nonlinear
mass-scale $M_{*}$ over the same redshift range -- while others have increased
their mass by factors of 200 to 300.

The spread in these trajectories increases substantially with decreasing
redshift. At $z=3$ the {\it rms} scatter in the log of the fractional growth 
is $0.168$, at $z=1$ it is $0.271$ and at $z=0$ it is $0.33$, slightly larger
than the initial separation in median mass between our SLQ and LLQ samples.
Although we do not display them, the mass accretion histories of haloes in other
cosmologies are very similar. This is, of course, expected.

Fig.~\ref{fig:traj} also shows the mass growth for the most massive (dotted)
and least massive (dashed) haloes in our SLQ sample.  By chance, the least
massive halo is one of the fastest growing, with a fractional growth rate
faster than that of the most massive halo at all redshifts. Indeed, this
``low-mass'' halo grows into a $2\times10^{15}\,\Msun$ object by $z=0$ ,
whereas the present-day descendant of the initially most massive halo is
actually slightly smaller ($1.8\times10^{15}\,\Msun$). Neither of these haloes
is in the extreme tail at $z=0$, ranking  2,224-th and 3,459-th in mass among
MXXL haloes, and being four times less massive than the most extreme
object. Curiously, none of the 20 most massive $z=0$ haloes in the MXXL has a
progenitor in the SLQ sample.

Finally, the thick solid line in Fig.~\ref{fig:traj} indicates the mass growth that
the most massive halo at $z=6$ would have to have in order to be the most
massive halo at all redshifts. This is close to the actual trajectory of the
largest halo until $z\sim3$, but at later times, other objects take over the
top spots.

\begin{figure} 
\includegraphics[width=8.5cm]{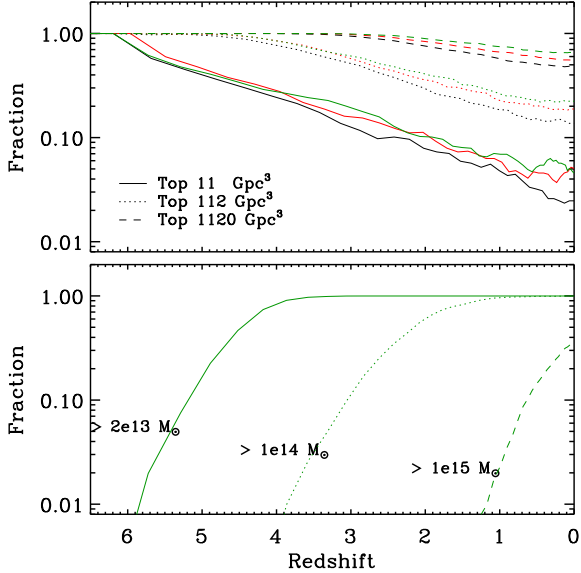} 
\caption{ Fraction of descendants of SLQ haloes that are among the 11 (solid),
  112 (dotted) or 1120 (dashed) most massive haloes per ${\rm Gpc}^3$
  at each later time (top panel), or that have a mass above
  $2.8\times10^{13}\,\Msun$, $1.4\times10^{14}\,\Msun$ or
  $1.4\times10^{15}\,\Msun$ (bottom panel). In the top panel, green,
  red and black lines correspond to the results for the M1, M7 and M3
  cosmologies.
\label{fig:rank}} \end{figure}

All these examples are not exclusive to extremely massive haloes at high
redshifts, but are a generic illustration that the most massive haloes at any
given epoch will no longer be the most massive haloes at a later time.  This is
a consequence of the diverse assembly histories of haloes in a hierarchical
universe.

We explore this behaviour directly in Fig.~\ref{fig:rank}. The top
panel indicates the fraction of SLQ halo descendants whose masses place them
above three abundance thresholds. The bottom panel shows a complementary
picture, showing the fraction of SLQ haloes that rise above various mass
thresholds at later times.

At redshift 3, all descendants of SLQ haloes have masses above
$2\times10^{13}\,\Msun$ and rank among the $3000$ most massive haloes per
cubic Gigaparsec. (Recall that initially these objects are {\it defined} as the
high-mass tail of the halo distribution with an abundance of 30 per cubic
Gpc.) In contrast, only $5\%$ are in haloes with $M > 1.4\times10^{14}\,\Msun$ and only
$20\%$ still rank among the $30$ most massive haloes per cubic Gigaparsec. With
time, the SLQ descendants fall further and further behind the most massive
haloes in the MXXL. By the present day, only $2-5\%$ (depending on cosmology) 
would still be included in a mass-limited sample with $n=11.6\,{\rm Gpc^{-3}}$ and 
about $50-70\%$ would be included in a sample with 100 times greater abundance.


\subsection{The environment of QSO haloes}

\begin{figure} 
\includegraphics[width=8.5cm]{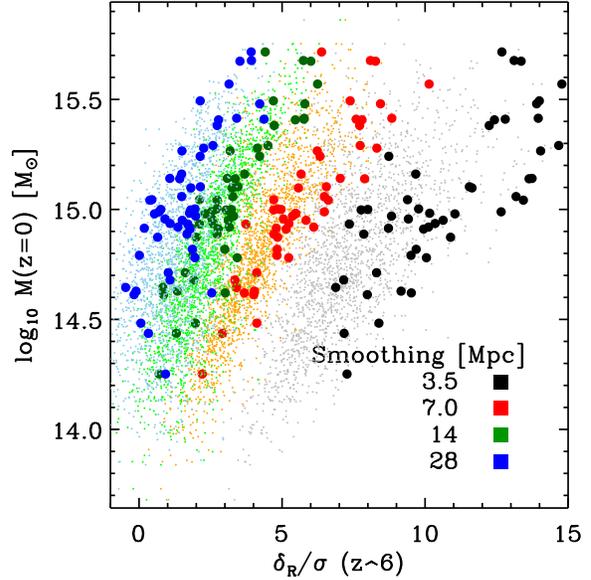} 
\caption{ Correlation between the overdensity in which a QSO halo sits at
  $z=6$ and the mass of its descendant at $z=0$.  The horizontal axis shows
  Gaussian-smoothed overdensity in units of its {\it rms} value $\sigma_{2.5}
  = 0.193$, $\sigma_{5} = 0.125$, $\sigma_{10} = 0.071$ and $\sigma_{20} =
  0.036$, where the suffix indicates the 1-D smoothing radius of the Gaussian
  in units of Mpc comoving. Different colours refer
  to different smoothing scales as indicated in the figure. Small and large
  points show the overdensities of the environments of SLQ and LLQ haloes,
  respectively.
\label{fig:over}} \end{figure}

Why do some high-redshift haloes keep growing rapidly until the present,
whereas others appear to shut down their accretion? Is this a random process
or can it be related to some halo property at $z\sim6$?
 
Fig~\ref{fig:over} shows how the masses of the $z=0$ descendants of our SLQ
and LLQ haloes correlate with their environment densities.  The overdensity
surrounding each high-redshift halo is computed on a variety of scales by
mapping the underlying dark matter distribution onto a $2048^3$ grid and then
convolving this density field with a Gaussian filter of different sizes.

Clearly, most of the QSO haloes live in overdense regions, but there is 
considerable diversity in how extreme these regions are. On small scales, all
live in at least a $3\sigma$ region, but $7\sigma$ is the typical
overdensity of an SLQ halo and $10\sigma$ that of an LLQ halo.  With increasing
smoothing, our quasar host haloes are found in progressively less extreme
regions and the separation between the SLQ and LLQ haloes decreases.
For a smoothing of $28$~Mpc the typical SLQ halo lives in a region
of overdensity $1.4\sigma$, and 10\% are located in regions of below-average 
density. Thus, our results suggest that it might be possible to find
a quasar in the middle of a $14-28$~Mpc underdense region, even
if quasars reside in the most massive haloes at $z\sim6$. 

Fig.~\ref{fig:over} also shows that there is clearly a strong correlation 
between the overdensity around a high-$z$ 
halo and the mass of its $z=0$ descendant. In fact, this correlation is
stronger than between the mass of the $z=6$ progenitor and that of its $z=0$
descendant (or with any other property in our catalogues).  For example, if we
consider the environment at $14$~Mpc, haloes that live in $< 1\sigma$
regions end up in haloes of $M \sim 2.8\times10^{14}\,\Msun$, whereas those
found in $>4\sigma$ regions typically end up in ten times more massive haloes
($2.8\times10^{15}\,\Msun$).  The scatter in descendant mass at a fixed
overdensity is $\sigma_{log M} = [0.061,\,0.044,\,0.045,\,0.063]$ for the
fields smoothed on $[3.5,\,7,\,14,\,28]$~Mpc scales, respectively. These
figures are to be compared with a scatter of $0.36$ for the sample as a
whole.  Thus, if the environment density surrounding a quasar is known, then
the mass of its $z=0$ descendant can be predicted $6-8$ times more accurately
than if it is not.

These results are easy to understand, since the masses of the $z=0$
descendants correspond to the mass contained within a sphere of radius about
$7-14$~Mpc, and an object can only form by $z=0$ if its material is already
overdense by a factor of about $1.68/D_{+}(z=6)$\footnote{$D_{+}(z)$ 
is the growth factor in units of its present-day value.}$=0.32$ at $z=6$. Our findings also
warn against a naive connection between objects at different redshifts -- the
linkage depends not only on the actual properties of the objects, but
also on their environment.

\subsection{Prior accretion histories}
\begin{figure} 
\includegraphics[width=8.5cm]{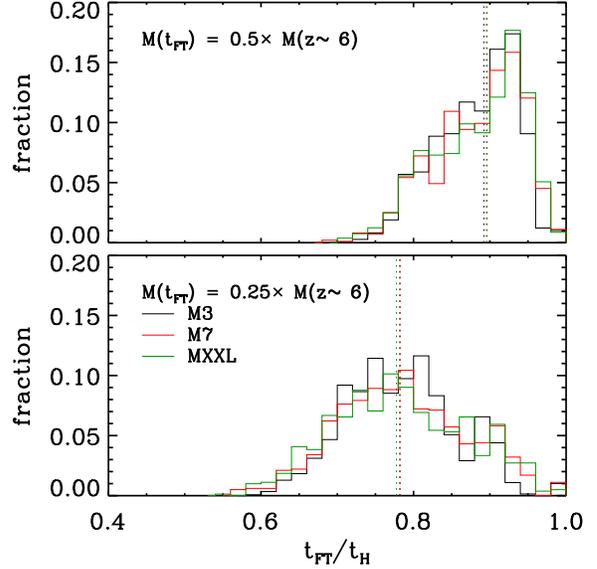} 
\caption{
The distribution of formation times for haloes in our SLQ sample. 
We define the formation time of a halo as the epoch when the main 
progenitor of a halo had $50\%$ or $25\%$ of the final halo mass, which we display
in the top and bottom panels respectively. These values are given in terms of
the age of the Universe at the time the SQL sample is identified and correspond
to $0.92$, $0.93$ and $0.94$ Gyr for the M1, M3 and M7 cosmologies, respectively.  
Dotted lines in both panels indicate the median values for each sample.
\label{fig:growth}} 
\end{figure}

Our $z\sim 6$ quasar host haloes are the most massive objects present at that
time and so, by definition, are the objects which had the highest mean mass
growth rates averaged over previous epochs. It is these extreme growth rates
which must supply the baryons needed to build up the supermassive black holes
and to fuel the extraordinarily luminous quasars which assume to lie in their
cores. In this final section, we examine when and how fast our LLQ and SLQ haloes
achieve their extreme masses.

We define an accretion history for each halo in our samples from the growth in
mass along the main branch of its assembly tree.  This branch is defined by
stepping back in time from the $z\sim 6$, object, selecting the most massive
progenitor of the current main branch object as the main branch object at the
immediately preceding step. Note that with this algorithm the main progenitor
at $z=10$, say, is not necessarily the most massive of all the $z=10$
progenitors. In fact, only about $\sim40\%$ of our SLQ haloes have an
identifiable main progenitor at $z\sim 10$ (i.e. with a mass above the
resolution limit, 20 particles or $1.2\times10^{11}\Msun$) even though the MXXL contains
125,000 identifiable haloes at this time. Conversely, of the $100$ most massive
MXXL haloes at $z=10$, only $30$ have a descendant among our SLQ sample at
$z\sim 6$. 

A consequence of these statistics is that SLQ haloes typically accrete most of
their mass in a relatively short period before they are identified.  This is
illustrated explicitly in Fig.~\ref{fig:growth}, which shows
histograms of the $25$ and $50\%$ growth times of haloes in each of our three SLQ
samples.  These are defined for each halo as the times since it had a quarter
and a half of its final mass, and they are given in units of the age of the
universe, $t_{\rm H}$, at the time the samples were defined. Median values are
$t_{\rm FT}/t_{\rm H} = 0.89$ and $0.78$, respectively for our two definitions of formation
time, and they are almost independent of the cosmological model. These values 
correspond to roughly $100$ and $200$ Myr prior $z\sim6$.

Recent accretion rates are clearly substantially larger than the mean value 
required to grow the halo in the Hubble time.  Median accretion rates for the 
last half of halo growth are $[9.7,\,8.9,\,7.6]\times10^{3}\,\Msun\,{\rm yr^{-1}}$ 
for the M1, M7 
and M3 samples, respectively. The growth times of our LLQ haloes are similarly
distributed to those shown in Fig.~\ref{fig:growth} and, as a result, their median
accretion rates are two to three times higher.  In all three cosmologies, the 
growth rates we find appear to be comfortably large enough to fuel even quasars 
as bright as the SDSS objects at $z\sim 6$, provided, of course that the
associated baryons are able to shed most of their angular momentum and reach
the central regions despite the tremendous luminosity being generated there.
 
\section{Conclusions}

In this paper we have combined the largest high-resolution cosmological
simulation to date with a scaling technique which allows a simulation to
represent structure growth in cosmologies other than that in which it was
originally carried out. This allows us to explore the properties and the
evolution of extremely massive haloes that might host $z\sim6$ quasars in three
cosmologies with parameters spanning the observationally allowed range.

We found significant differences in the growth of such haloes subsequent to
their identification at $z\sim 6$.  Some increase their mass by a factor of
$200$ by $z=0$ whereas others grow only by a factor of $10$. As a result, the
descendants of bright high-redshift quasars are inferred to live in haloes with
a wide range of halo masses today.  The median descendant mass of haloes in a
mass-limited sample with space density $11.6\,{\rm Gpc}^{-3}$ at $z\sim6$ is
$5.7\times10^{14}\,\Msun$ for a WMAP7-like cosmology, while more massive objects
with 30 times lower abundance, thus matching the directly observed number
density of luminous SDSS quasars, end up in haloes with median mass about a
factor of two higher. In both samples descendants spread in mass by a factor of
several above and below this median. Conversely, in this same cosmology, only
$4\%$ of present-day haloes with mass above $2.8\times10^{15}\,\Msun$,
corresponding to space density $11.6\,{\rm Gpc}^{-3}$, have a progenitor at $z\sim
6$ with mass above $7.1\times10^{12}\,\Msun$ and so would be considered a
potential quasar host at this same abundance.  These figures change only
slightly for the other two cosmologies we consider.

Another aspect of the same effect is that haloes ranked among the most massive
at a given time, will gradually occupy lower positions and other haloes,
initially less massive, will take over the top positions.  The
dissimilar mass growth is also expected to influence the galaxies that would
form in these haloes: two haloes of the same mass may thus host 
galaxies with different properties \citep{Zhu2006}. Since the large-scale clustering
of haloes at given mass depends on assembly history, this violates the
core  assumption of HOD modelling and simple abundance matching techniques,
namely that the galaxy population in a halo depends only on its
mass, not on its large-scale environment. 

We find that the best way to predict the later growth of $z=6$ haloes 
is to look at their local environment on $14$~Mpc scales, which correlates with 
the halo mass at $z=0$
much more strongly than the actual mass at $z\sim6$.  We emphasise that the
behaviour we describe in the paper is not restricted to $z\sim6$ haloes, but it
is a general feature expected in hierarchical growth, where the initial
amplitudes of different Fourier modes are independent of each other.  This
behaviour is a example of a general property of certain mathematical
distributions known as ``regression to the mean'', which describes the
migration of an extreme sample to a less extreme one at a later 
time.\footnote{This phenomenon was first pointed out by Sir Francis Galton, a cousin
of Charles Darwin, in the 19th century.} The ``regression to the mean'' can only be avoided if
there were a perfect and monotonically increasing relation between the mass of
a halo and that of its descendant.  In this case the rank order of haloes by
mass is perfectly preserved. Thus, extreme haloes at, for instance, $z=6$ beget
equally extreme haloes at $z=0$. However, we have confirmed the expectations
of earlier results \citep{deLucia2007,Trenti2008, Overzier2009} that this 
situation does not apply in hierarchical structure formation from Gaussian initial
conditions.

Finally, we explored the assembly of QSO haloes prior to their identification
at $z\sim6$, finding one of the fastest accretion rates ever seen in simulated
objects: a median of about $8.6\times10^{3}\,\Msun\,{\rm yr^{-1}}$ (but up to a
factor 2 larger) for the 100~Myrs preceding identification at $z\sim6$, almost
independent of cosmological model. This appears sufficient to fuel the bright 
quasars observed in the SDSS.  
 
\section*{Acknowledgments}

We would like to thank the referee for constructive comments. RA
acknowledges useful discussions with Silvia Bonoli, Roderik Overzier and
Francesco Shankar.  We thank the staff at the J\"ulich Supercomputer Centre in
Germany for their technical assistance which helped us to successfully complete
the MXXL simulation.  The initial conditions software was developed and tested
on COSMA-4 which is part of the DiRAC Facility jointly funded by STFC, the
Large Facilities Capital Fund of BIS, and Durham University. RA and SW are
supported by Advanced Grant 246797 ``GALFORMOD'' from the European Research
Council. VS and SW acknowledge support by the DFG Collaborative Research
Network TR33 ``The Dark Universe''.  CSF acknowledges a Royal Society Wolfson
Research Merit award and ERC Advanced Investigator grant ``COSMIWAY''. This
work was supported in part by an STFC rolling grant to the ICC.

\bibliographystyle{mn2e} \bibliography{quas}

\label{lastpage} \end{document}